\documentstyle[aps,preprint]{revtex}

\begin{document}
\draft
\title{Introducing Small-World Network Effect to Critical Dynamics}
\author{Jian-Yang Zhu$^{1,2}$ and Han Zhu$^3$}
\address{$^1$CCAST (World Laboratory), Box 8730, Beijing 100080, China\\
$^2$Department of Physics, Beijing Normal University, Beijing 100875, China%
\thanks{%
Mailing address.}\\
$^3$Department of Physics, Nanjing University, Nanjing, 210093, China}
\maketitle

\begin{abstract}
We analytically investigate the kinetic Gaussian model and the
one-dimensional kinetic Ising model on two typical small-world networks
(SWN), the adding-type and the rewiring-type. The general approaches and
some basic equations are systematically formulated. The rigorous
investigation of the Glauber-type kinetic Gaussian model shows the
mean-field-like global influence on the dynamic evolution of the individual
spins. Accordingly a simplified method is presented and tested, and believed
to be a good choice for the mean-field transition widely (in fact, without
exception so far) observed on SWN. It yields the evolving equation of the
Kawasaki-type Gaussian model. In the one-dimensional Ising model, the $p$%
-dependence of the critical point is analytically obtained and the
inexistence of such a threshold $p_c$, for a finite temperature transition,
is confirmed. The static critical exponents, $\gamma $ and $\beta $ are in
accordance with the results of the recent Monte Carlo simulations, and also
with the mean-field critical behavior of the system. We also prove that the
SWN effect does not change the dynamic critical exponent, $z=2$, for this
model. The observed influence of the long-range randomness on the critical
point indicates two obviously different hidden mechanisms.
\end{abstract}

\pacs{PACS number(s): 89.75.-k, 64.60.Ht, 64.60.Cn, 64.60.Fr}

\section{Introduction}

Small world networks (SWN) are those intermediate between a regular lattice
and a random graph (see \cite{SWNinitial,book,meanfieldsolution,Ising1} and
references therein for review). They are believed to catch the essence of
networks in reality, such as neural networks\cite{neural}, power grids,
social networks, and documents on World Wide Web\cite{social1,social2},
where remote vertices, while locally clustered, often have the chance to be
connected via shortcuts. Since 1998, when Watts and Strogatz presented a
simple model showing SWN effects\cite{SWNinitial}, it has been studied
intensively and extensively\cite{variety}. From the point of view of
statistical physics, the presence of shortcuts assists the system to behave
as a whole, showing global coherence and new, SWN behavior, possibly apart
from their ordinary properties people are familiar with.

One may be curious about, to what extent the features of phase transitions
will be different in spin-lattice models built on small-world networks.
Because the SWN effect widely exists in reality, this question is also of
much significance. Although new and interesting features have been revealed
recently\cite{Ising1,Ising2,Ising3,Ising4,Ising5,Ising6,Ising7,XY,XY2}, it
is still far from being completely answered. The dynamic aspect has been
even less well understood\cite{randomwalk}. Naturally we may expect the
evolution of a single spin to be influenced partly by the overall system,
however we find it difficult to offer more specific information. This is due
to the complexity of the dynamics itself, and the often formidable
mathematical task.

In this article, we report our work in this interesting problem ---the
critical dynamics of spin-lattice models built on SWN. As the introductory
content we shall first discuss the general approach. Then, among the various
model systems, we choose two special ones for a detailed investigation: the
Gaussian model, which is relatively easy in mathematics, and the
one-dimensional Ising model. New and interesting kinetic features are
revealed analytically and the study in the dynamic aspect also yields much
information about the static properties, such as the critical point.

This article is organized as follows: In Sec. \ref{Sec. 2} we give the
definition of spin-lattice models built on SWN. Sec. \ref{Sec. 3} contains
the discussion of the general approach (along with a brief review of the
dynamic mechanisms). The direct application on the kinetic Gaussian model
can be found in Sec. \ref{Sec. 4}. In Sec. \ref{Sec. 5}, we present a
simplified method and prove its validity in Sec. \ref{Sec. 5.1} by comparing
its result with the rigorous one obtained in \ref{Sec. 4}. Sec. \ref{Sec.
5.2} and Sec. \ref{Sec. 5.3} are devoted to the further applications of this
method on the Kawasaki-type Gaussian model and the one-dimensional
Glauber-type Ising model, respectively. In Sec. \ref{Sec. 6}, the influence
of the randomness on the critical point is analyzed. Sec. \ref{Sec. 7} is
the summarization with some discussions.

\section{Spin-lattice models built on small-world networks}

\label{Sec. 2}

Following the first prototype\cite{SWNinitial} of SWN, there have been a
number of variants (see, for example, Ref. \cite{variant}) in two basic
groups, which can be constructed as the following: the initial network is,
for example, a one dimensional loop of $N$ vertices, each vertex being
connected to its $2k$ nearest neighbors. (1) Each pair of random vertices
are additionally connected with probability $p_A$; (2) the vertices are then
visited one after the other, and each link connecting a vertex to one of its 
$k$ nearest neighbors in the clock wise sense is left in place with
probability $1-p_R$, and with probability $p_R$ is reconnected to a randomly
chosen other vertex. We may call the first group adding-type small-world
networks (A-SWN) and the second group rewiring-type (R-SWN). Both of these
modifications introduce long range connections. The above algorithms can be
extended to systems with higher dimensionality and even fractal structures.
(For example, for a spin located somewhere in a cubic lattice, $3$ of the $6$
bonds in some fixed directions will be reconnected with probability $p_R$
respectively.) These constructions, with any (infinitesimally low) fraction
of shortcuts, allow to reconcile local properties of a regular network
(clustering effect) with global properties of a random one (the average
distance $l\sim \log _{10}N$).

A spin-lattice model built on SWN, which is the focus of this article, can
be thus defined: In a $D$-dimensional regular network consisting of $N$
spins with periodic boundary condition, each spin is linked to its $2kD$
nearest neighbors (in this article we will choose $k=1$). Then, (1) a
certain number of supplemental links are added; (2) a portion of the bonds
are rewired. Whether it is in group (1) (A-SWN) or group (2) (R-SWN), two
spins connected by a shortcut act in the same way as those connected by a
regular bond, that is, their interaction contributes to the system
Hamiltonian and they obey the redistribution (exchange) mechanism. In short,
there is no difference between a long range bond and a regular one. Other
definitions may exist. For example, because these small-world networks are
model systems for the networks in reality, one may have reason to choose
certain physical quantities, such as $J$ in the Ising model, to be
different. However, although it is not necessary, we will limit the scope of
this article to the above rule.

\section{The Dynamic Mechanism}

\label{Sec. 3}

The various dynamic processes in the critical phenomena are believed to be
governed by two basic mechanisms, Glauber-type with order parameter
nonconserved and Kawasaki-type with order parameter conserved. Recently, a
Glauber-type single-spin transition mechanism\cite{sst1,sst2} and a
Kawasaki-type spin-pair redistribution mechanism\cite{ssr,ssrr} have been
presented as the natural generalizations of Glauber's flipping mechanism\cite
{Glauber} and Kawasaki's exchange mechanism\cite{Kawasaki}, respectively.
They are generally applicable and mathematically well organized. Our work
begins with a review of these two mechanisms. It is for your better
understanding of the calculations in Sec. \ref{Sec. 4} and \ref{Sec. 5}, and
it also might be a convenient reference for further studies.

\subsection{Single-spin transition mechanism}

\label{Sec. 3.1}

Glauber's single-spin flipping mechanism allows an Ising system to evolve
with its spins flipping to their opposite. In the single-spin transition
mechanism\cite{sst1}, a single spin $\sigma _i$ may change itself to any
possible values, $\hat{\sigma}_i$, and the master equation is 
\begin{equation}
\frac d{dt}P(\{\sigma \},t)=-\sum_i\sum_{\hat{\sigma}_i}\left[ W_i(\sigma
_i\rightarrow \hat{\sigma}_i)P(\{\sigma \},t)-W_i(\hat{\sigma}_i\rightarrow
\sigma _i)P(\{\sigma _{j\neq i}\},\hat{\sigma}_i,t)\right] .
\label{sst-master}
\end{equation}
The transition probability is in a normalized form determined by a heat
Boltzmann factor, 
\begin{equation}
W_i(\sigma _i\rightarrow \hat{\sigma}_i)=\frac 1{Q_i}\exp \left[ -\beta 
{\cal H}\left( \left\{ \sigma _{j\neq i}\right\} ,\hat{\sigma}_i\right)
\right] ,\text{ }Q_i=\sum_{\hat{\sigma}_i}\exp \left[ -\beta {\cal H}\left(
\left\{ \sigma _{j\neq i}\right\} ,\hat{\sigma}_i\right) \right] ,
\label{sst-w}
\end{equation}
where $\beta =1/k_BT$. Based on the master equation, Eq.(\ref{sst-master}),
one can prove that, 
\begin{equation}
\frac{dq_k\left( t\right) }{dt}=-q_k\left( t\right) +\sum_{\left\{ \sigma
\right\} }\left[ \sum_{\hat{\sigma}_k}\hat{\sigma}_kW_k\left( \sigma
_k\rightarrow \hat{\sigma}_k\right) \right] P\left( \left\{ \sigma \right\}
;t\right) ,  \label{G-q}
\end{equation}
where $q_k\left( t\right) \equiv \sum_{\left\{ \sigma \right\} }\sigma
_kP\left( \left\{ \sigma \right\} ;t\right) $. There are also corresponding
equations for the correlation functions.

\subsection{Spin-pair redistribution mechanism}

\label{Sec. 3.2}

Kawasaki's spin-pair exchange mechanism allows an Ising system to evolve
with its nearest neighbors exchanging their spin values. In the spin-pair
redistribution mechanism\cite{ssr}, two connected spins, $\sigma _j$ and $%
\sigma _l$, may change to any possible values, $\hat{\sigma}_j$ and $\hat{%
\sigma}_l$, as long as their sum are conserved. The master equation is 
\begin{eqnarray}
\frac d{dt}P(\{\sigma \},t) &=&\sum_{\left\langle jl\right\rangle }\sum_{%
\hat{\sigma}_j,\hat{\sigma}_l}\left[ -W_{jl}\left( \sigma _j\sigma
_l\rightarrow \hat{\sigma}_j\hat{\sigma}_l\right) P\left( \left\{ \sigma
\right\} ;t\right) \right.  \nonumber \\
&&\left. +W_{jl}\left( \hat{\sigma}_j\hat{\sigma}_l\rightarrow \sigma
_j\sigma _l\right) P\left( \left\{ \sigma _{i\neq j},\sigma _{l\neq
k}\right\} ,\hat{\sigma}_j,\hat{\sigma}_l;t\right) \right] .
\label{K-master}
\end{eqnarray}
The redistribution probability is also in a normalized form determined by a
heat Boltzmann factor, 
\begin{equation}
W_{jl}\left( \sigma _j\sigma _l\rightarrow \hat{\sigma}_j\hat{\sigma}%
_l\right) =\frac 1{Q_{jl}}\delta _{\sigma _j+\sigma _l,\hat{\sigma}_j+\hat{%
\sigma}_l}\exp \left[ -\beta {\cal H}\left( \left\{ \sigma _m\right\}
_{m\neq j,l},\hat{\sigma}_j,\hat{\sigma}_l\right) \right] ,  \label{K-w}
\end{equation}
where the normalization factor $Q_{jl}$ is 
\[
Q_{jl}=\sum_{\hat{\sigma}_j,\hat{\sigma}_l}\delta _{\sigma _j+\sigma _l,\hat{%
\sigma}_j+\hat{\sigma}_l}\exp \left[ -\beta {\cal H}\left( \left\{ \sigma
_m\right\} _{m\neq j,l},\hat{\sigma}_j,\hat{\sigma}_l\right) \right] . 
\]
Based on the master equation, Eq. (\ref{K-master}), one can prove that, 
\begin{equation}
\frac{dq_k\left( t\right) }{dt}=-2Dq_k\left( t\right) +\sum_{\left\{ \sigma
\right\} }\sum_w\left[ \sum_{\hat{\sigma}_k,\hat{\sigma}_{k+w}}\hat{\sigma}%
_kW_{k,k+w}\left( \sigma _k\sigma _{k+w}\rightarrow \hat{\sigma}_k\hat{\sigma%
}_{k+w}\right) \right] P\left( \left\{ \sigma \right\} ;t\right) ,
\label{K-q}
\end{equation}
where $D$ is the dimensionality and $\sum_w$ stands for the summation taken
over the nearest neighbors.

\subsection{How to apply them on SWN}

\label{Sec. 3.3}

In the construction of SWN, according to a certain probability, we will have
a whole set of possible realizations. So the theoretically correct way of
treating dynamic systems built on SWN actually consists of three steps:
First we have to make a full list of all the possible realizations and point
out the probability of each one of them. Second, we treat each system
respectively (apply the dynamic mechanism and obtain the master equation and
the physical quantities of interest). Third, we give the expectation value
with all these results. This is cumbersome, but conceptually
straightforward. In Sec. \ref{Sec. 5} we shall discuss whether there is any
simplified method. (In fact, there is.)

\section{Kinetic Gaussian model governed by the Glauber-type mechanism}

\label{Sec. 4}

We have discussed the general approach right above in Sec. \ref{Sec. 3.3},
and in this section we will directly perform the calculations according to
the three steps, to study the $3$D kinetic Gaussian model governed by the
Glauber-type single-spin transition mechanism (see Sec. \ref{Sec. 3.1}). The
calculation to be carried out is very long but fortunately we can borrow
some results from our earlier studies\cite{sst1,sst2}.

The Gaussian model, proposed by Berlin and Kac, is a continuous-spin model.
Its Hamiltonian, 
\begin{equation}
-\beta {\cal H}=K\sum_{\left\langle i,j\right\rangle }\sigma _i\sigma _j,
\label{G-Ham}
\end{equation}
where $K=J/k_BT$. The spins $\sigma _k$ can take any real value between $%
\left( -\infty ,+\infty \right) $, and the probability of finding a given
spin between $\sigma _k$ and $\sigma _k+d\sigma _k$ is assumed to be the
Gaussian-type distribution, $f\left( \sigma _k\right) d\sigma _k\sim \exp
\left( -\frac b2\sigma _k^2\right) d\sigma _k$, where $b$ is a distribution
constant independent of temperature. Thus, the summation for the spin value
turns into the integration $\sum_\sigma \rightarrow \int_{-\infty }^\infty
f\left( \sigma \right) d\sigma $. This model has been studied often as a
starting point for investigations of other systems.

Governed by the Glauber-type single-spin transition mechanism, the
expectation value of single spin obeys\cite{sst1} 
\begin{equation}
\frac{dq_k\left( t\right) }{dt}=-q_k\left( t\right) +\frac Kb%
\sum_{\left\langle q_{k^{\prime }},q_k\right\rangle }q_{k^{\prime }}\left(
t\right) .  \label{G-q-result}
\end{equation}
Since on SWN a spin may have a neighbor located very far from it, in the
following calculations the summation is to include every spin that is
connected with $\sigma _k$. Taking average of Eq. (\ref{G-q-result}), we can
obtain the evolution of the magnetization $M\left( t\right) =\sum_kq_k\left(
t\right) /N$.

\subsection{kinetic Gaussian model on an adding-type small-world network}

\label{Sec. 4.2}

First we treat the $3$D kinetic Glauber-type Gaussian model on A-SWN
consisting of $N$ (a very large number) spins. The Kawasaki-type is still
tractable, but the mathematical task will be more complex. We will leave
that till Sec. \ref{Sec. 5}. Besides the regular bonds, each pair of spins,
no matter how far apart, is connected via an additional bond with
probability $p_A$. Actually there are $2^{N\left( N-1\right) /2}$ different
networks, each with a given probability. As a model system for the networks
in reality, we expect the number of these bonds, $n\sim N\left( N-1\right)
p_A/2$, to be much smaller than $N$. So practically we require $p_AN\ll 1$.
With respect to a specific spin $\sigma _{ijk}$, all the networks can be
divided into $N$ groups listed as the following.

($0$). There is no random bond on $\sigma _{ijk}$, and the probability is $%
\left( 1-p_A\right) ^{N-1}$. According to Eq.(\ref{G-q-result}) 
\[
\frac d{dt}q_{ijk}\left( t\right) =-q_{ijk}\left( t\right) +\frac Kb%
\sum_w\left( q_{i+w,j,k}\left( t\right) +q_{i,j+w,k}\left( t\right)
+q_{i,j,k+w}\left( t\right) \right) . 
\]

($1$). There is only one random bond on $\sigma _{ijk}$, and the probability
is $C_{N-1}^1p_A\left( 1-p_A\right) ^{N-2}$. In fact this group can be
further divided into $N-1$ subgroups, each corresponding to a specific spin
connected to $\sigma _{ijk}$, and each with the same probability $p_A\left(
1-p_A\right) ^{N-2}$. Averaging them we get 
\[
\frac{dq_{ijk}\left( t\right) }{dt}=-q_{ijk}\left( t\right) +\frac Kb%
\sum_w\left( q_{i+w,j,k}\left( t\right) +q_{i,j+w,k}\left( t\right)
+q_{i,j,k+w}\left( t\right) \right) +\frac KbM\left( t\right) . 
\]

$\cdots \cdots $.

($n$). There are $n$ random bonds on $\sigma _{ijk}$, and the probability is 
$C_{N-1}^np_A^n\left( 1-p_A\right) ^{N-n-1}$. Similarly, in average we get, 
\[
\frac{dq_{ijk}\left( t\right) }{dt}=-q_{ijk}\left( t\right) +\frac Kb%
\sum_w\left( q_{i+w,j,k}\left( t\right) +q_{i,j+w,k}\left( t\right)
+q_{i,j,k+w}\left( t\right) \right) +\frac{nK}bM\left( t\right) . 
\]

$\cdots \cdots $.

($N-1$). There are $N-1$ random bonds on $\sigma _{ijk}$, and the
probability is $p_A^{N-1}.$%
\[
\frac{dq_{ijk}\left( t\right) }{dt}=-q_{ijk}\left( t\right) +\frac Kb%
\sum_w\left( q_{i+w,j,k}\left( t\right) +q_{i,j+w,k}\left( t\right)
+q_{i,j,k+w}\left( t\right) \right) +\frac Kb\left( N-1\right) M\left(
t\right) . 
\]

Thus, over all the realizations, 
\begin{eqnarray}
\frac{dq_{ijk}\left( t\right) }{dt} &=&-q_{ijk}\left( t\right) +\frac Kb%
\sum_w\left( q_{i+w,j,k}\left( t\right) +q_{i,j+w,k}\left( t\right)
+q_{i,j,k+w}\left( t\right) \right)  \nonumber \\
&&+\frac Kb\sum_{n=0}^{N-1}C_{N-1}^np_A^n\left( 1-p_A\right) ^{N-n-1}nM.
\end{eqnarray}
Fortunately we find the following relationship 
\begin{equation}
\sum_{n=0}^{N-1}C_{N-1}^np_A^n\left( 1-p_A\right) ^{N-n-1}n=\left(
N-1\right) p_A,  \label{important-relationship}
\end{equation}
and thus 
\begin{equation}
\frac{dq_{ijk}\left( t\right) }{dt}=-q_{ijk}\left( t\right) +\frac Kb%
\sum_w\left( q_{i+w,j,k}\left( t\right) +q_{i,j+w,k}\left( t\right)
+q_{i,j,k+w}\left( t\right) \right) +\frac Kb\left( N-1\right) p_AM\left(
t\right) ,  \label{q-gaussian-A-g-3}
\end{equation}
Similar results can be found in one and two dimensional models, and taking
average we obtain 
\begin{equation}
\frac{dM\left( t\right) }{dt}=-\left[ 1-\frac{2DK}b-\frac Kb\left(
N-1\right) p_A\right] M\left( t\right) ,  \label{M-gaussian-A-g}
\end{equation}
where $D$ is the dimensionality. The solution of Eq.(\ref{M-gaussian-A-g})
is 
\[
M\left( t\right) =M\left( 0\right) \exp \left( -t/\tau \right) , 
\]
where the relaxation time 
\[
\tau =\frac 1{1-K/K_c^A}, 
\]
and the critical point 
\begin{equation}
K_c^A=\frac b{2D+\left( N-1\right) p_A}.
\end{equation}
It is well known that the critical point of the regular model $%
K_c^{reg}=b/2D $, we can clearly see that {\it the critical temperature will
get higher as more long range bonds are added.} Actually, for a vertex
located on an A-SWN, the number of the long range bonds $n_A\sim \left(
N-1\right) p_A$. In the small-world region, where the expected $n_A$ for
most of the vertices is very small, the change of the critical point will be
almost unperceivable. (The analysis of this result can be found in Sec. \ref
{Sec. 6}.)

\subsection{kinetic Gaussian model on a rewiring-type small-world network}

\label{Sec. 4.3}

Second we treat the kinetic Gaussian model on a R-SWN with characteristic
probability $p_R$. Because of the length, here we only give the details of
the $1$D case. With regards to a specific spin $\sigma _k$, all the networks
can be divided into four major groups listed as the following, ($1$) both of
the two regular bonds on $\sigma _k$, connecting $\sigma _{k-1}$ and $\sigma
_{k+1}$, are not rewired; ($2$) the bond connecting $\sigma _{k-1}$ and $%
\sigma _k$ is left unchanged but that connecting $\sigma _k$ and $\sigma
_{k+1}$ is rewired; ($3$) the bond connecting $\sigma _k$ and $\sigma _{k+1}$
is left unchanged but that connecting $\sigma _{k-1}$ and $\sigma _k$ is
rewired; ($4$) both of them are rewired. The final result comes from the
summation of all the four parts (see Appendix \ref{Appendix A} for the
details): 
\begin{equation}
\frac{dq_k\left( t\right) }{dt}=-q_k\left( t\right) +p_R\frac{2K}bM\left(
t\right) +\left( 1-p_R\right) \frac Kb\left( q_{k-1}\left( t\right)
+q_{k+1}\left( t\right) \right) ,  \label{q-gaussian-R-1}
\end{equation}
In two and three dimensional models, there are $2^4$ and $2^6$ major groups
respectively, and they can be treated in the same way. Similar results are
obtained: 
\begin{equation}
\frac{dM\left( t\right) }{dt}=-\left( 1-\frac{2DK}b\right) M\left( t\right) .
\label{M-gaussian-R-any}
\end{equation}
Obviously the relaxation time 
\[
\tau =\frac 1{1-K/K_c^R}, 
\]
where the critical point 
\begin{equation}
K_c^R=K_c^{reg}=b/2D.
\end{equation}

To summarize, we strictly obtain the evolution of the Gaussian model built
on SWN. On A-SWN, the critical temperature will get higher as more long
range bonds are added, while on R-SWN the critical temperature is unchanged.
With the dynamic scaling hypothesis $\tau \sim \xi ^z\sim \left|
T-T_c\right| ^{-z\nu }$, we have $z\nu =1$, where $z$ is the dynamic
critical exponent and $\nu $ is the correlation length critical exponent. We
shall leave the summarization of its properties to Sec. \ref{Sec. 7}. The
influence of the long range bonds on the critical point, which is a rather
interesting topic, will be discussed in Sec. \ref{Sec. 6}.

The Gaussian model, being an idealization, often has the value of serving as
a starting point for the more general studies. The evolving equations of the
individual spins show distinctly the influence of the global coherence, which%
{\it \ automatically takes a mean-field-like form as if it comes from an
averaged spin.} This character, as well as some already-proved facts in
earlier studies about the static behavior, provides us with some helpful
hints for the simplification of the method, which is the topic of the
following section.

\section{The Simplified Method}

\label{Sec. 5}

Presently, for the critical dynamics on SWN, we still lack a
well-established approach, which should be both theoretically reliable and
practically feasible. The mentioned-above method (and the results) is
theoretically rigorous, but it is too complex for the other models, even the
simplest one-dimensional Ising model. In the above section, we have pointed
out that, in {\it the dynamic evolution} of the Gaussian model built on SWN,
the influence of the system as a whole on individual spins is mean-field
like. On the other hand, recent studies on the Ising model and the $XY$
model built on SWN have also shown that the phase transition is of {\it the
mean-field-type} (see \cite{Ising1,Ising2,Ising3,Ising4,Ising5,XY} and Sec. 
\ref{Sec. 5.3} for details). The various model systems built on SWN probably
belong to the same mean-field universality class, and they might be treated
with the same dynamic approach. In this problem, which is found to be of the
mean-field nature, the following method may be the best choice. We deem all
the possible networks as a single one. The effective Hamiltonian of a
spin-lattice model built on such a network is defined as the expectation
value over all possible realizations. Its effective behavior, e.g. the
redistribution between two vertices connected with each other, is also the
averaged result. For example, two sites, $\sigma _i$ and $\sigma _j$, are
connected with probability $p$, then redistribution occurs between them with
probability $pW_{ij}$. (It is the basic assumption when we are treating a
system governed by the Kawasaki-type mechanism). Then we can directly apply
the dynamic mechanisms to this system. This method, which is of the
mean-field nature, is certainly more tractable in mathematics. We apply it
to the kinetic Gaussian model again and see if it will lead to the same
result.

\subsection{Application in kinetic Gaussian model governed by the
Glauber-type mechanism}

\label{Sec. 5.1}

In the $3$D Gaussian model ($N$ spins in total) built on A-SWN with periodic
boundary condition, the effective Hamiltonian 
\begin{eqnarray}
-\beta {\cal H} &=&K\sum_{i,j,k}\sigma _{ijk}\left( \sigma _{i+1,j,k}+\sigma
_{i,j+1,k}+\sigma _{i,j,k+1}\right)   \nonumber \\
&&+\frac 12Kp_A\sum_{i,j,k}\sigma _{ijk}\sum_{i^{\prime },j^{\prime
},k^{\prime }}\sigma _{i^{\prime }j^{\prime }k^{\prime }}-\frac 12%
Kp_A\sum_{i,j,k}\sigma _{ijk}^2.
\end{eqnarray}
With the Glauber-type transition mechanism, we begin to derive the
single-spin evolving equation according to Eq.(\ref{G-q}), along the same
line of the calculations in Ref.\cite{sst1}. We obtain 
\begin{equation}
\sum_{\hat{\sigma}_{ijk}}\hat{\sigma}_{ijk}W_{ijk}\left( \sigma
_{ijk}\rightarrow \hat{\sigma}_{ijk}\right) =\frac Kb\left[ \sum_w\left(
\sigma _{i+w,jk}+\sigma _{ij+w,k}+\sigma _{ij,k+w}\right) +\left( N-1\right)
p_A\bar{\sigma}\right] ,  \label{sim-G-combined}
\end{equation}
where 
\[
\bar{\sigma}=\frac 1{N-1}\left( \sum_{lmn}\sigma _{lmn}-\sigma _{ijk}\right)
\approx \frac 1N\sum_{lmn}\sigma _{lmn}.
\]
Substituting Eq. (\ref{sim-G-combined}) into Eq. (\ref{G-q}) we get 
\begin{equation}
\frac{dq_{ijk}\left( t\right) }{dt}=-q_{ijk}\left( t\right) +\frac Kb%
\sum_{w=\pm 1}\left( q_{i+w,jk}\left( t\right) +q_{ij+w,k}\left( t\right)
+q_{ij,k+w}\left( t\right) \right) +\frac Kb\left( N-1\right) p_AM\left(
t\right) .
\end{equation}
Similar results can be obtained for one and two dimensional models. These
results, along with those for $M\left( t\right) $, are exactly in accordance
with what we have obtained in Sec. \ref{Sec. 3.2}, Eqs. (\ref
{q-gaussian-A-g-3}) and (\ref{M-gaussian-A-g}).

In the $3$D kinetic Gaussian model built on R-SWN, the effective
Hamiltonian, 
\begin{eqnarray}
-\beta {\cal H} &=&K\left( 1-p_R\right) \sum_{i,j,k}\sigma _{ijk}\left(
\sigma _{i+1,jk}+\sigma _{i,j+1,k}+\sigma _{i,j,k+1}\right)  \nonumber \\
&&+Kp_R\frac 3N\sum_{i,j,k}\sigma _{ijk}\sum_{i^{\prime },j^{\prime
},k^{\prime }}\sigma _{i^{\prime }j^{\prime }k^{\prime }}-Kp_R\frac 3N%
\sum_{i,j,k}\sigma _{ijk}^2.  \label{eff-Ham-R-3}
\end{eqnarray}
Similar calculations yield 
\[
\frac{dq_{ijk}}{dt}=-q_{ijk}+p_R\frac{6K}bM+\left( 1-p_R\right) \frac Kb%
\sum_w\left( q_{i+w,j,k}+q_{i,j+w,k}+q_{i,j,k+w}\right) . 
\]
This is also in accordance with the rigorous result, Eqs. (\ref
{q-gaussian-R-1},\ref{M-gaussian-R-any}). Thus, in the kinetic Gaussian
model this simplified method yields the same results as the rigorous ones
obtained with the more complex standard method. As various spin-lattice
models, such as the Ising model, are believed to show mean-field behavior on
SWN, we believe that this simplified method is able to provide at least
qualitatively correct information. In this sense, it is very different from
the mean-field approximations taken in other universality classes. In later
studies, we will use it to study some more complex problems. The following
are two examples.

\subsection{Application in kinetic Gaussian model governed by a
Kawasaki-type mechanism}

\label{Sec. 5.2}Now we apply this simplified method to study the diffusion
process in the kinetic Gaussian model built on SWN. Although we still have
to deal with many complex equations, it is relatively easy compared with the
formidable task of the standard approach. In such processes, the system is
governed by the Kawasaki-type redistribution mechanism. As already mentioned
at the beginning of this section, the system behavior, just as the
Hamiltonian, is also averaged over all possible realizations. The basic
equations for the Kawasaki-type dynamics listed below are generally
applicable in various order-parameter-conserved processes.

($1$) On $D$-dimensional A-SWN: Accordingly the master equation should be
modified as, 
\begin{eqnarray}
\frac d{dt}P(\{\sigma \},t) &=&\sum_{\left\langle jl\right\rangle }\sum_{%
\hat{\sigma}_j,\hat{\sigma}_l}\left[ -W_{jl}\left( \sigma _j\sigma
_l\rightarrow \hat{\sigma}_j\hat{\sigma}_l\right) P\left( \left\{ \sigma
\right\} ;t\right) \right.  \nonumber \\
&&\left. +W_{jl}\left( \hat{\sigma}_j\hat{\sigma}_l\rightarrow \sigma
_j\sigma _l\right) P\left( \left\{ \sigma _{i\neq j},\sigma _{l\neq
k}\right\} ,\hat{\sigma}_j,\hat{\sigma}_l;t\right) \right]  \nonumber \\
&&+\frac 12p_A\sum_j\sum_{l\neq j}\sum_{\hat{\sigma}_j,\hat{\sigma}_l}\left[
-W_{jl}\left( \sigma _j\sigma _l\rightarrow \hat{\sigma}_j\hat{\sigma}%
_l\right) P\left( \left\{ \sigma \right\} ;t\right) \right.  \nonumber \\
&&\left. +W_{jl}\left( \hat{\sigma}_j\hat{\sigma}_l\rightarrow \sigma
_j\sigma _l\right) P\left( \left\{ \sigma _{i\neq j},\sigma _{l\neq
k}\right\} ,\hat{\sigma}_j,\hat{\sigma}_l;t\right) \right] ,
\label{sim-K-master}
\end{eqnarray}
where the redistribution probability $W_{jl}\left( \sigma _j\sigma
_l\rightarrow \hat{\sigma}_j\hat{\sigma}_l\right) $ is of the same form as
Eq. (\ref{K-w}). With Eq. (\ref{sim-K-master}) we can get that 
\begin{eqnarray}
\frac{dq_k\left( t\right) }{dt} &=&-2Dq_k\left( t\right) +\sum_{\left\{
\sigma \right\} }\sum_w\left[ \sum_{\hat{\sigma}_k,\hat{\sigma}_{k+w}}\hat{%
\sigma}_kW_{k,k+w}\left( \sigma _k\sigma _{k+w}\rightarrow \hat{\sigma}_k%
\hat{\sigma}_{k+w}\right) \right] P\left( \left\{ \sigma \right\} ;t\right) 
\nonumber \\
&&+p_A\left\{ -\left( N-1\right) q_k\left( t\right) +\sum_{\left\{ \sigma
\right\} }\left[ \sum_{l\neq k}\sum_{\hat{\sigma}_k,\hat{\sigma}_l}\hat{%
\sigma}_kW_{kl}\left( \sigma _k\sigma _l\rightarrow \hat{\sigma}_k\hat{\sigma%
}_l\right) \right] P\left( \left\{ \sigma \right\} ;t\right) \right\} 
\nonumber \\
&\equiv &A_k^{\left( 1\right) }+p_AA_k^{\left( 2\right) }.
\label{sim-K-Gaussian-A}
\end{eqnarray}

($2$) On $D$-dimensional R-SWN: Accordingly the master equation should be
modified as, 
\begin{eqnarray}
\frac d{dt}P(\{\sigma \},t) &=&\left( 1-p_R\right) \sum_{\left\langle
jl\right\rangle }\sum_{\hat{\sigma}_j,\hat{\sigma}_l}\left[ -W_{jl}\left(
\sigma _j\sigma _l\rightarrow \hat{\sigma}_j\hat{\sigma}_l\right) P\left(
\left\{ \sigma \right\} ;t\right) \right.   \nonumber \\
&&\left. +W_{jl}\left( \hat{\sigma}_j\hat{\sigma}_l\rightarrow \sigma
_j\sigma _l\right) P\left( \left\{ \sigma _{i\neq j},\sigma _{l\neq
k}\right\} ,\hat{\sigma}_j,\hat{\sigma}_l;t\right) \right]   \nonumber \\
&&+Dp_R\frac 1{N-1}\sum_j\sum_{l\neq j}\sum_{\hat{\sigma}_j,\hat{\sigma}%
_l}\left[ -W_{jl}\left( \sigma _j\sigma _l\rightarrow \hat{\sigma}_j\hat{%
\sigma}_l\right) P\left( \left\{ \sigma \right\} ;t\right) \right.  
\nonumber \\
&&\left. +W_{jl}\left( \hat{\sigma}_j\hat{\sigma}_l\rightarrow \sigma
_j\sigma _l\right) P\left( \left\{ \sigma _{i\neq j},\sigma _{l\neq
k}\right\} ,\hat{\sigma}_j,\hat{\sigma}_l;t\right) \right] .
\end{eqnarray}
The redistribution probability $W_{jl}\left( \sigma _j\sigma _l\rightarrow 
\hat{\sigma}_j\hat{\sigma}_l\right) $ is of the same form as Eq.(\ref{K-w},
and 
\begin{eqnarray}
\frac{dq_k\left( t\right) }{dt} &=&\left( 1-p_R\right) \left\{ -2Dq_k\left(
t\right) +\sum_{\left\{ \sigma \right\} }\sum_w\left[ \sum_{\hat{\sigma}_k,%
\hat{\sigma}_{k+w}}\hat{\sigma}_kW_{k,k+w}\left( \sigma _k\sigma
_{k+w}\rightarrow \hat{\sigma}_k\hat{\sigma}_{k+w}\right) \right] P\left(
\left\{ \sigma \right\} ;t\right) \right\}   \nonumber \\
&&+\frac{Dp_R}{N-1}\left\{ -\left( N-1\right) q_k\left( t\right)
+\sum_{\left\{ \sigma \right\} }\left[ \sum_{l\neq k}\sum_{\hat{\sigma}_k,%
\hat{\sigma}_l}\hat{\sigma}_kW_{kl}\left( \sigma _k\sigma _l\rightarrow \hat{%
\sigma}_k\hat{\sigma}_l\right) \right] P\left( \left\{ \sigma \right\}
;t\right) \right\}   \nonumber \\
&\equiv &\left( 1-p_R\right) R_k^{\left( 1\right) }+\frac{Dp_R}{N-1}%
R_k^{\left( 2\right) }.  \label{sim-K-Gaussian-R}
\end{eqnarray}
Although $A_k^{\left( 1,2\right) }$ and $R_k^{\left( 1,2\right) }$ are of
the same form respectively, we use different symbols because they are
actually different (determined by the Hamiltonian-dependent redistribution
probability $W$).

First we treat the Gaussian model built on A-SWN. The first part, $%
A_{ijk}^{\left( 1\right) }$, comes from the redistribution between the
nearest neighbors. In the $3$D case, with Eq. (\ref{K-w}) we can obtain 
\begin{eqnarray}
A_{ijk}^{\left( 1\right) } &=&\frac 1{2\left( b+K\right) }b\left\{ \left[
\left( q_{i+1,j,k}-q_{ijk}\right) -\left( q_{ijk}-q_{i-1,j,k}\right) \right]
\right.  \nonumber \\
&&\left. +\left[ \left( q_{i,j+1,k}-q_{ijk}\right) -\left(
q_{ijk}-q_{i,j-1,k}\right) \right] +\left[ \left( q_{i,j,k+1}-q_{ijk}\right)
-\left( q_{ijk}-q_{i,j,k-1}\right) \right] \right\}  \nonumber \\
&&+\frac K{2\left( b+K\right) }\left[ 2\left(
2q_{i-1,j,k}-q_{i-1,j+1,k}-q_{i-1,j-1,k}\right) +\left(
2q_{i-1,j,k}-q_{ijk}-q_{i-2,j,k}\right) \right.  \nonumber \\
&&+2\left( 2q_{i+1,j,k}-q_{i+1,j+1,k}-q_{i+1,j-1,k}\right) +\left(
2q_{i+1,j,k}-q_{ijk}-q_{i+2,j,k}\right)  \nonumber \\
&&+2\left( 2q_{i,j-1,k}-q_{i,j-1,k+1}-q_{i,j-1,k-1}\right) +\left(
2q_{i,j-1,k}-q_{ijk}-q_{i,j-2,k}\right)  \nonumber \\
&&+2\left( 2q_{i,j+1,k}-q_{i,j+1,k+1}-q_{i,j+1,k-1}\right) +\left(
2q_{i,j+1,k}-q_{ijk}-q_{i,j+2,k}\right)  \nonumber \\
&&+2\left( 2q_{i,j,k-1}-q_{i+1,j,k-1}-q_{i-1,j,k-1}\right) +\left(
2q_{i,j,k-1}-q_{ijk}-q_{i,j,k-2}\right)  \nonumber \\
&&\left. +2\left( 2q_{i,j,k+1}-q_{i+1,j,k+1}-q_{i-1,j,k+1}\right) +\left(
2q_{i,j,k+1}-q_{ijk}-q_{i,j,k+2}\right) \right] .  \label{3d-Q-result}
\end{eqnarray}
Actually this is not as complex as it seems. With the lattice constant $a$
we can transform the above expression to be 
\begin{equation}
A_{ijk}^{\left( 1\right) }=\frac{3a^2}{b+K}\left( \frac b6-K\right) \nabla
^2q\left( {\bf r,}t\right) .  \label{A1}
\end{equation}

The second part, $A_{ijk}^{\left( 2\right) }$, comes from the redistribution
between $\sigma _{ijk}$ and the farther spins, $\sigma _{i^{\prime
}j^{\prime }k^{\prime }}$. The result is 
\begin{eqnarray}
A_{ijk}^{\left( 2\right) } &=&-\left( N-1\right) q_{ijk}\left( t\right) +%
\frac 12\sum_{i^{\prime }j^{\prime }k^{\prime }\neq ijk}\left( q_{ijk}\left(
t\right) +q_{i^{\prime }j^{\prime }k^{\prime }}\left( t\right) \right) 
\nonumber \\
&&+\frac K{2b}\sum_{i^{\prime }j^{\prime }k^{\prime }\neq ijk}\sum_w\left[
\left( q_{i+w,j,k}\left( t\right) +q_{i,j+w,k}\left( t\right)
+q_{i,j,k+w}\left( t\right) \right) \right.  \nonumber \\
&&\left. -\left( q_{i^{\prime }+w,j^{\prime },k^{\prime }}\left( t\right)
+q_{i^{\prime },j^{\prime }+w,k^{\prime }}\left( t\right) +q_{i^{\prime
},j^{\prime },k^{\prime }+w}\left( t\right) \right) \right] ,  \label{A2}
\end{eqnarray}
Substituting Eqs.(\ref{A1}) and (\ref{A2}) into Eq.(\ref{sim-K-Gaussian-A}),
we obtain the evolving equation of the Gaussian model built on $3$D A-SWN, 
\begin{eqnarray}
\frac{\partial q\left( {\bf r},t\right) }{\partial t} &=&\left[ \frac{3a^2}{%
b+K}\left( \frac b6-K\right) +p_A\left( N-1\right) \frac{a^2K}{2b}\right]
\nabla ^2q\left( {\bf r,}t\right)  \nonumber \\
&&+\frac{p_A\left( N-1\right) }2\left( 1-\frac{6K}b\right) \left( M\left(
t\right) -q\left( {\bf r,}t\right) \right) .  \label{sim-Gaussian-K-A-result}
\end{eqnarray}

Similarly, on $3$D R-SWN we obtain 
\begin{eqnarray}
\frac{\partial q\left( {\bf r},t\right) }{\partial t} &=&\left\{ \left(
1-p_R\right) \frac{3a^2}{b+K\left( 1-p_R\right) }\left[ \frac b6-K\left(
1-p_R\right) \right] +3p_R\frac K{2b}\left( 1-p_R\right) \right\} \nabla
^2q\left( {\bf r,}t\right)  \nonumber \\
&&+\frac{3p_R}2\left[ 1-\frac{6K}b\left( 1-p_R\right) \right] \left( M\left(
t\right) -q\left( {\bf r,}t\right) \right) .  \label{sim-Gaussian-K-R-result}
\end{eqnarray}

From these two equations we can clearly see the influence of the system
built on SWN as a whole on individual spins. On regular lattices the
evolution of the system can be explained by the diffusion mechanism, while
on SWN, to some degree, the individual spins will automatically adjust
itself to approach the average magnetization. This is obviously the result
of the global coherence introduced by the small fraction of the long range
bonds.

On regular lattice, where the evolution is pure diffusion, $\partial q\left( 
{\bf r},t\right) /\partial t={\cal D}\nabla ^2q\left( {\bf r,}t\right) $,
and the diffusion coefficient ${\cal D}$ will vanish near the critical
point. However, on A-SWN, where $\partial q\left( {\bf r},t\right) /\partial
t={\cal D}^{\prime }\nabla ^2q\left( {\bf r,}t\right) +{\cal C}\left[
M\left( t\right) -q\left( {\bf r,}t\right) \right] $, two temperatures, $%
T_{c1}^A$ and $T_{c2}^A$ can be obtained by setting ${\cal D}^{\prime }$ and 
${\cal C}$ to be zero, respectively. $T_{c1}^A$ will be lower than the
critical temperature on a regular lattice, $T_c^{reg}$, and suggests that
the point at which the diffusion stops will be lowered by the randomness. $%
T_{c2}^A$ equals $T_c^{reg}$, and at this point the evolution will be pure
diffusion. Similarly there are also two temperatures for R-SWN, but we will
have $T_{c1}^R<T_{c2}^R<T_c^{reg}$.

Obviously, the system behavior strongly depends on the temperature. For
example, we study a one-dimensional system and the initial magnetization is $%
q\left( x,0\right) =\sin x$. When the temperature $T>T_{c2}$, both ${\cal D}%
^{\prime }$ and ${\cal C}$ are positive, and the magnetization will approach
homogeneity. When $T<T_{c1}$, both ${\cal D}^{\prime }$ and ${\cal C}$ are
negative, and the inhomogeneity will be getting more remarkable during the
evolution. When $T_{c2}<T<T_{c1}$, ${\cal D}^{\prime }>0$ but ${\cal C}<0$,
and this is a more complex region. Although here we can still easily predict
the system behavior with Eq. (\ref{sim-Gaussian-K-A-result}) and obtain a
stationary point, generally the evolution will be strongly dependent on the
local magnetization.

\subsection{Application in Ising model}

\label{Sec. 5.3}

The second application of the simplified method is on the one-dimensional
ferromagnetic Ising model governed by the Glauber-type mechanism. Recently
it has been studied both analytically\cite{Ising1,Ising2} and with Monte
Carlo simulations\cite{Ising3,Ising4,Ising5,Ising6,Ising7}. Due to the
mathematical difficulties, the study was not going very smoothly at the
beginning. In Ref. \cite{Ising2}, Gitterman concluded that the random
long-range interactions, the number of which above a minimal value, lead to%
{\it \ a phase transition}. In Ref. \cite{Ising1}, Barrat and Weigt used
some approximations and expected a finite critical point ''at least for
sufficiently large $p$ and $k\geq 2$'' on R-SWN. They found out that {\it %
this transition is of the mean-field type}. Although they could not
calculate the transition analytically, the numerical computation
demonstrated {\it a nonvanishing order parameter in the presence of a
vanishingly small fraction of shortcuts}, for $k=2$ and $k=3$. The numerical
results seemed to support the following relationship, $T_c\sim -2k/\log
_{10}p_R$. The more recent Monte Carlo studies have proved the
above-mentioned conclusions on A-SWN and R-SWN respectively, but in those
cases $k=1$. The analysis of the relationship between $T_c$ and $p_R$ can be
found in Ref. \cite{Ising4}. The critical exponents obtained from the
simulation\cite{Ising4,Ising5,XY}, $\beta \approx 1/2$, $\alpha \approx 0$,
and $\nu \approx 1/2$, further establish the mean-field character. However,
although there is substantial numerical proof, the inexistence of such a
threshold $p_c$, for a finite temperature transition, is still to be
confirmed.

In this article, we will apply the simplified method, which is of the
mean-field nature, to study the dynamic properties of the one-dimensional
Ising model built on both A-SWN and R-SWN. We hope the success in the
Gaussian model will continue in Ising system, of which the behavior is
already known to be the mean-field-type. Although we can not obtain the full
picture of the evolution, we are able to get the exact $p$-dependence of the
critical point, and some interesting critical exponents. As will be shown
below, our result agrees perfectly with the above-mentioned numerical
simulation.

We find that the system shows very similar behavior on A-SWN and R-SWN. We
shall give the details of R-SWN only, and report the results of A-SWN later.

On a R-SWN, the effective Hamiltonian is the same as Eq. (\ref{eff-Ham-R-3})
(the one-dimensional version). We substitute it into the single-spin
evolving equation, Eq.(\ref{G-q}) and obtain, 
\begin{equation}
\frac{dq_k\left( t\right) }{dt}=-q_k\left( t\right) +\sum_{\left\{ \sigma
\right\} }\tanh \left[ K\left( 1-p_R\right) \left( \sigma _{k-1}+\sigma
_{k+1}\right) +2Kp_R\bar{\sigma}\right] P\left( \left\{ \sigma \right\}
;t\right) ,  \label{Ising-R-q}
\end{equation}
where $\bar{\sigma}=\sum_{j\neq k}\sigma _j/\left( N-1\right) \approx
\sum_j\sigma _j/N$.

If $p_R=0$, then it is the one-dimensional Ising model on regular lattice,
which we are familiar with. One can continue to write 
\[
\frac{dq_k\left( t\right) }{dt}=-q_k\left( t\right) +\frac 12\left(
q_{k-1}+q_{k+1}\right) \tanh 2K, 
\]
and 
\[
\frac{dM\left( t\right) }{dt}=-M\left( t\right) \left( 1-\tanh 2K\right) . 
\]
It yields $M\left( t\right) \propto e^{-t/\tau }$, where $\tau =\left(
1-\tanh 2K\right) ^{-1}$. When $K\rightarrow K_c^{reg}=\infty $, $\tau \sim
\xi ^z\sim \left( e^{2K}\right) ^2$, and thus $z=2$.

If the rewiring probability $p_R=1$, then 
\[
\frac{dq_k\left( t\right) }{dt}=-q_k\left( t\right) +\sum_{\left\{ \sigma
\right\} }\tanh \left( 2K\bar{\sigma}\right) P\left( \left\{ \sigma \right\}
;t\right) , 
\]
and 
\[
\frac{dM\left( t\right) }{dt}=-M\left( t\right) +\sum_{\left\{ \sigma
\right\} }\tanh \left( 2K\bar{\sigma}\right) P\left( \left\{ \sigma \right\}
;t\right) . 
\]
Although $\left\langle \tanh \left( 2K\bar{\sigma}\right) \right\rangle \neq
\tanh \left( 2KM\right) $, if we study the case when, near the critical
point, the system is in almost thorough disorder, we will have $\left\langle
\tanh \left( 2K\bar{\sigma}\right) \right\rangle \sim \tanh \left(
2KM\right) \sim 2KM$, and 
\[
\frac{dM\left( t\right) }{dt}\sim -M\left( t\right) +\tanh \left[ 2KM\left(
t\right) \right] . 
\]
This helps to determine the critical point, $K_c=1/2$. When $K<K_c=1/2$,
then the system will be stable in a disordered state with $M=0$, but when $%
K\geq K_c=1/2$, there appears some kind of order. Taking Taylor expansion,
one can find that when $K$ is near $K_c$, $M\sim \left( K-K_c\right) ^{1/2}$%
. This leads to $\beta =1/2$ (in this specific situation). We also get the
relaxation time $\tau \sim \left| K-K_c\right| ^{-1}$, with the scaling
hypothesis (see below) we have $z\nu =1$.

Qualitatively similar situation should also be found when $p_R$ is between $%
0 $ and $1$. First we assume that there is a critical temperature, above
which the system is disordered and below which there begins to show nonzero
magnetization. From Eq.(\ref{Ising-R-q}) we find that if initially $M=0$ is
given, then the system will stay in this disordered state. But below the
critical temperature this equilibrium will not be stable. We can determine
the critical point introducing a small perturbation. When $M\rightarrow 0$, 
\begin{eqnarray}
\frac{dq_k\left( t\right) }{dt} &\approx &-q_k\left( t\right) +\sum_{\left\{
\sigma \right\} }\tanh \left[ K\left( 1-p_R\right) \left( \sigma
_{k-1}+\sigma _{k+1}\right) \right] P\left( \left\{ \sigma \right\} ;t\right)
\nonumber \\
&&+2Kp_R\sum_{\left\{ \sigma \right\} }\bar{\sigma}\left\{ 1-\tanh ^2\left[
K\left( 1-p_R\right) \left( \sigma _{k-1}+\sigma _{k+1}\right) \right]
\right\} P\left( \left\{ \sigma \right\} ;t\right)  \nonumber \\
&=&-q_k\left( t\right) +\frac 12\left( q_{k-1}+q_{k+1}\right) \tanh \left[
2K\left( 1-p_R\right) \right]  \nonumber \\
&&+2Kp_R\sum_{\left\{ \sigma \right\} }\bar{\sigma}\left\{ 1-\frac 12\left(
1+\sigma _{k-1}\sigma _{k+1}\right) \tanh ^2\left[ 2K\left( 1-p_R\right)
\right] \right\} P\left( \left\{ \sigma \right\} ;t\right) .
\label{dqdtIsing}
\end{eqnarray}
Because $M$ is very small, we believe $\frac 1N\sum_k\left\langle \bar{\sigma%
}\sigma _{k-1}\sigma _{k+1}\right\rangle $ is an even smaller quantity of
higher order. Thus 
\begin{eqnarray}
\frac{dM\left( t\right) }{dt} &=&-M+M\tanh \left[ 2K\left( 1-p_R\right)
\right]  \nonumber \\
&&+2Kp_R\left\{ 1-\frac 12\tanh ^2\left[ 2K\left( 1-p_R\right) \right]
\right\} M.  \label{Ising-M}
\end{eqnarray}
The critical point can be determined as 
\begin{equation}
\tanh \left[ 2K_c^R\left( 1-p_R\right) \right] +2K_c^Rp_R\left\{ 1-\frac 12%
\tanh ^2\left[ 2K_c^R\left( 1-p_R\right) \right] \right\} =1.
\label{Ising-Kc-R}
\end{equation}
If $K<K_c^R$, the disordered state $M=0$ will be stable, but if $K>K_c^R$, a
small perturbation will drive the system apart from the disordered state
towards nonzero magnetization. Now we continue to find several interesting
critical exponents.

($1$) $\chi \sim \left| T-T_c\right| ^{-\gamma }$: Near the critical point, $%
M_{eq}\rightarrow 0$. If a weak field $H$ is introduced, we will have 
\[
-\beta {\cal H}=\sum_k\sigma _k\left[ K\left( 1-p_R\right) \sigma _{k+1}+p_R%
\frac K{N-1}\sum_{j\neq k}\sigma _j+\frac H{k_BT}\right] , 
\]
and 
\begin{equation}
\frac{dq_k\left( t\right) }{dt}=-q_k\left( t\right) +\sum_{\left\{ \sigma
\right\} }\tanh \left[ K\left( 1-p_R\right) \left( \sigma _{k-1}+\sigma
_{k+1}\right) +2Kp_R\bar{\sigma}+\frac H{k_BT}\right] P\left( \left\{ \sigma
\right\} ;t\right) .
\end{equation}
Following the same way as that taken in the calculation of the critical
point (in fact, we can just replace $2Kp_R\bar{\sigma}$ by $2Kp_R\bar{\sigma}%
+H/k_BT$), we get 
\begin{eqnarray}
\frac d{dt}M\left( t\right) &=&-M\left( t\right) +M\left( t\right) \tanh
\left[ 2K\left( 1-p_R\right) \right]  \nonumber \\
&&+\left( 2Kp_RM\left( t\right) +\frac H{k_BT}\right) \left\{ 1-\frac 12%
\tanh ^2\left[ 2K\left( 1-p_R\right) \right] \right\} .  \label{beta-2}
\end{eqnarray}
From Eq.(\ref{beta-2}) one can easily find that, in a system in equilibrium
near the critical point, $K=K_c^R+\Delta $, and 
\begin{equation}
\chi \equiv \frac{\partial M}{\partial H}\sim \Delta ^{-1}.
\end{equation}
Thus, $\gamma =1$.

($2$) $M\sim \left| T-T_c\right| ^\beta $: When $M\rightarrow 0$, we take 
\[
\frac 1N\sum_k\left\langle \bar{\sigma}\sigma _{k-1}\sigma
_{k+1}\right\rangle \sim M^3. 
\]
Then from Eq. (\ref{dqdtIsing}) we can get 
\begin{eqnarray*}
\frac{dM\left( t\right) }{dt} &\approx &-\left( 1-\tanh \left[ 2K\left(
1-p_R\right) \right] -2Kp_R\left\{ 1-\frac 12\tanh ^2\left[ 2K\left(
1-p_R\right) \right] \right\} \right) M\left( t\right) \\
&&-Kp_R\tanh ^2\left[ 2K\left( 1-p_R\right) \right] M^3\left( t\right) .
\end{eqnarray*}
When $K-K_c^R=\Delta \rightarrow 0$, let $dM\left( t\right) /dt=0$, and one
will get $M^2\sim \Delta $ by taking Taylor expansion. Thus, the critical
exponent $\beta =1/2$.

($3$) $\tau \sim \xi ^z\sim \left| T-T_c\right| ^{-z\nu }$: When studying
the critical slowing down, we can assume $M$ is very small and thus use Eq.(%
\ref{Ising-M}). It yields 
\[
M\left( t\right) =M\left( 0\right) e^{-t/\tau }, 
\]
where 
\[
\tau ^{-1}=1-\tanh \left[ 2K\left( 1-p_R\right) \right] -2Kp_R\left\{ 1-%
\frac 12\tanh ^2\left[ 2K\left( 1-p_R\right) \right] \right\} . 
\]
If $K\rightarrow K_c^R$, then $\tau \rightarrow \dot{\infty}$. If $%
K=K_c^R-\Delta $, and $\Delta \rightarrow 0$, then one will find $\tau \sim
\Delta ^{-1}$. Thus $z\nu =1$. It has been found in Monte Carlo simulations
that $\nu \approx 1/2$, so $z=2$. It is of the same value as that obtained
on the regular lattice.

$1$D Ising model on A-SWN show {\it qualitatively the same} behavior: Its
critical point can be determined as 
\begin{equation}
\tanh 2K_c^A+K_c^A\left( N-1\right) p_A\left( 1-\frac 12\tanh
^22K_c^A\right) =1,  \label{Ising-Kc-A}
\end{equation}
and we have found {\it the same critical exponents, }$\gamma $, $\beta $ and 
$z$.

For a vertex on a A-SWN lattice, the number of the long range bonds is $%
n_A\sim \left( N-1\right) p_A$, while for a vertex on a R-SWN lattice, $%
n_R\sim 2p_R$. The $n$-dependence of critical point $K_c$ of $1$D Ising
model on A-SWN and R-SW, Eqs. (\ref{Ising-Kc-R}) and (\ref{Ising-Kc-A}), can
be found in Fig. 1. One can clearly see the mentioned-above approximate
relationship, $K_c^A\propto -\log _{10}n_A\sim -\log _{10}\left[ \left(
N-1\right) p_A\right] $, or $K_c^R\propto -\log _{10}n_R\sim -\log
_{10}\left( 2p_R\right) $.

When $n_A$ and $n_R$ are small enough (in the small world region), from Eqs.
(\ref{Ising-Kc-R})(\ref{Ising-Kc-A}) we can get

\[
n_R\simeq \left( 1-\tanh 2K_c^R\right) /\left( \frac 12K_c^R\tanh
^22K_c^R\right) , 
\]
\[
n_A\simeq \left( 1-\tanh 2K_c^A\right) /\left[ K_c^A\left( 1-\frac 12\tanh
^22K_c^A\right) \right] . 
\]
When $n_A$ and $n_R$ are approaching zero, $K_c^{A,R}\rightarrow \infty $.
For the same value of the critical point, $\left( n_R-n_A\right) \rightarrow
0^{+}$. As shown in Fig. 1, for most of the region, the two curves are very
close to each other. However, though the difference may be infinitesimal,
the curve of $n_R$ is always above that of $n_A$, as is distinct when they
are relatively large.

\section{The influence of the randomness on the critical point}

\label{Sec. 6}

On the behavior of the critical point, our results show interesting contrast
between the Gaussian model and the Ising model. Here, we shall mention
another interesting model system, the $D$-dimensional mean-field (MF) Ising
model, which might help us understand this problem. For a randomly selected
spin in this model, each of its $2D$ nearest neighbors is replaced by an
averaged one, and it is well known that the critical point is $K_c=1/2D$. A
simple calculation will yield that on R-SWN $K_c^R$ will not change, while
on A-SWN, since the long range bonds increase the contact of a spin with the
system, $K_c^A=1/\left( 2D+n_A\right) $.

Our result of the $1$D Ising model is in contrast to that of the MF Ising
model and the Gaussian model, while for each one of them the critical
temperature will be very close on A-SWN and R-SWN (the former will be
higher), in the small-world region. This may be a result of the totally
different role played by the long range bonds. ($1a$) In the $D$-dimensional
MF Ising model, the critical point is solely determined by the mean
coordination number, which decides the coupling between an individual spin
and the system. It is unchanged on R-SWN but will increase on A-SWN. Thus,
the critical temperature will stay unaltered on R-SWN but will be slightly
increased by the long range bonds on A-SWN, which typically take up only a
small fraction. ($1b$) As is shown by earlier studies\cite{sst1}, the
Gaussian model, though being a very different system, has a critical point
which also {\it only depends on the mean coordination number}. On SWN, its
critical point is very similar to the MF Ising model. It is believed to be
mainly a mathematical result, and to understand this we shall review the
calculations in Sec. \ref{Sec. 4}. On R-SWN, for an individual spin, the
long range interaction partly replaces the nearest-neighbor (n.n) coupling.
However, in geography, the coordination number can be considered unchanged,
since no bonds are created or eliminated (they are just redirected). As a
result, on the right-hand side of the evolving equations of the spins, the
lost part of the n.n\ coupling is exactly compensated by the MF term. The
critical point is determined by taking average of the evolving equations of
the spins, which only consist of linear terms. Thus, it is mathematically
straightforward that the critical point will stay the same. On A-SWN, the
consideration is similar, except that here the coordinate number will be
slightly increased. ($2$) The cross-over observed in the one-dimensional
Ising model is certainly governed by a different mechanism. For example, on
R-SWN, there are two competing length scales: the correlation length $\xi
\sim \exp \left( 2J/k_BT\right) $, and the characteristic length scale of
the SWN, which can be taken as the typical distance between the ends of a
shortcut\cite{distance}, $\zeta \sim p_R^{-1/D}=p_R^{-1}$. When $\xi <\zeta $%
, the system basically behaves as a regular lattice, otherwise it will show
MF behavior as an effect of the long range interactions. The transition
occurs at $\xi \approx \zeta $, suggesting a critical point $T_c\sim \left|
\log _{10}p_R\right| ^{-1}$. Since the two structures, R-SWN and A-SWN, have
the same length scales, the critical point can hardly be separated in the
small-world region. However, interestingly we find that the mechanism in ($1$%
), i.e., long-range bonds $\rightarrow $ coordination number (interaction
energy) $\rightarrow $ critical point, can also be observed, though being a
minor factor. As is mentioned in Sec. \ref{Sec. 5.3}, for the same expected
number of long range bonds, the critical temperature on A-SWN will be higher
than that obtained on R-SWN. We expect it to be a general phenomenon, though
not yet reported by the numerical simulations.

\section{Summary and Discussions}

\label{Sec. 7}

In this article, we study the critical dynamics, Glauber-type and
Kawasaki-type, on two typical small-world networks, adding-type (A-SWN) and
rewiring type (R-SWN).

{\it The logical sequence}: As the introductory content we discuss the
general approach of the critical dynamics, which is theoretically
straightforward but may be mathematically too complex. We directly apply it
to the kinetic Gaussian model governed by Glauber-type mechanism and obtain
its evolution. We observe that, in the {\it dynamic} evolution of the
individual spins, the influence of the system as a whole, which is the
result of the presence of the long range bonds, takes the mean-field-like
form as if it comes from an averaged spin. At the same time, earlier studies
have revealed the mean-field (MF) {\it static} behavior of the Ising model
and $XY$ model. These both suggest us to present the following simplified
method. All the SWN realizations are deemed as a single one, with both the
effective Hamiltonian and the effective behavior averaged over all of them.
It is tested in the same model and exactly leads to the same rigorous
result. Then this method, which is believed to be theoretically reliable and
mathematically feasible, is applied to two more difficult problems, the
Gaussian model governed by Kawasaki-type mechanism and the one-dimensional
kinetic Ising model.

{\it The Gaussian model:} ($1$) Whether it is built on A-SWN or R-SWN, the
long range bonds introduce the mean-field-like global influence of the
system to the dynamic evolution of the individual spins. On A-SWN such
influence is additional while on R-SWN it partly replaces that of the
neighboring spins. ($2$) On A-SWN, as more long range bonds are added, the
critical temperature gets higher; but on R-SWN it does not differ at all
from that of the regular lattice. This interesting discrepancy is explained
in Sec. \ref{Sec. 6}. ($3$) In both networks, the relaxation time, $\tau
=1/\left( 1-K/K_c\right) $, and thus $z\nu =1$. This dynamic property has
also been obtained on regular lattices and fractal lattices\cite
{sst1,sst2,ssr,ssrr}. It is highly universal, independent of the geometric
structure and the dynamic mechanism. ($4$) On SWN, the evolution of the
Kawasaki-type model can be viewed as the combination of two mechanisms, the
diffusion, and the automatic adjustment of the single spin to approach the
average magnetization. The pure diffusion equation, $\frac \partial {%
\partial t}q\left( {\bf r},t\right) ={\cal D}\nabla ^2q\left( {\bf r}%
,t\right) $, will be modified as, $\frac \partial {\partial t}q\left( {\bf r}%
,t\right) ={\cal D}^{\prime }\nabla ^2q\left( {\bf r},t\right) +C\left(
M\left( t\right) -q\left( {\bf r},t\right) \right) $. By setting ${\cal D}%
^{\prime }$ and ${\cal C}$ to be zero we will get two competing
characteristic temperatures, instead of the single definition of the
critical point for the regular lattices (${\cal D}=0$). The temperature
dependence of the dynamic evolution is discussed in Sec. \ref{Sec. 5.2}.

{\it The Ising model }analytically studied by the simplified method: The
system shows very similar behavior on the two networks, A-SWN and R-SWN.
Introducing very small perturbation of the local magnetization to the
disordered state, we obtain the critical point by judging the stability of
the equilibrium. The inexistence of such a threshold $p_c$ for a finite
temperature transition is confirmed\footnote{%
As mentioned above, for some physical considerations one may define a
different $J^{\prime }$ for the long range bonds, which might be much
smaller than $J$. With this method we can prove there is not such a
threshold $J_c^{\prime }$ either.}. From the dynamic equation we obtain the
critical exponents $\gamma \ $and $\beta $ in agreement with the numerical
simulation and the already-proved MF behavior of the system. The relaxation
time is divergent near the critical point as $\tau \sim \left| T-T_c\right|
^{-1}$ and thus $z\nu =1$ (note the same relationship in the Gaussian
model). In the $1$D Ising model, the SWN effect does not change the dynamic
critical exponent $z=2$.

{\it The influence of the randomness on the critical point}: Our result of
the $1$D Ising model is in contrast to that of the MF Ising model and the
Gaussian model. For each one of them the critical temperature on A-SWN, $%
T_c^A$, will be higher than that on R-SWN, $T_c^R$, obtained for the same
expected number of the long range bonds, though the difference may be hardly
perceivable in the small-world region. A detailed analysis of the
responsible mechanisms can be found in Sec. \ref{Sec. 6}.

{\it Prospects: }We hope the further studies of critical dynamics will
continue to reveal interesting dynamic characteristics in the widely
existing critical phenomena combined with SWN. The simplified method shall
become a useful tool in this field. As the phase transition is of the MF
nature, this method is certainly the best choice, and in this sense it is
basically different from the custom MF approximation applied to other
universality classes. Presently besides the numerical study, analytical
treatment in the dynamic aspect is scarce compared to the study of the
static properties\cite{Ising1,Ising2}. In fact, at least in some cases, such
a study may be feasible and fruitful indeed. Our work, especially that on
the Ising model, also shows that the study of the dynamic aspect is often
able to yield much information of the general properties, in a relatively
convenient way.

\acknowledgments

This work was supported by the National Natural Science Foundation of China
under Grant No. 10075025.

\appendix

\section{$3$D Gaussian model on R-SWN}

\label{Appendix A}

($1$). The two regular bonds on $\sigma _k$ (connecting $\sigma _{k-1}$ and $%
\sigma _{k+1}$) have not been rewired, and the probability is $\left(
1-p\right) ^2$. This group can be further divided into many subgroups:

($1.0$). There are no random bond on $\sigma _k$. The probability is $\left(
1-p\right) ^2\left( 1-\frac 1{N-1}p\right) ^{N-2}$. 
\[
\frac{dq_k}{dt}=-q_k+\frac Kb\left( q_{k-1}+q_{k+1}\right) . 
\]

($1.1$). There is only one random bond on $\sigma _k$. The probability is $%
\left( 1-p\right) ^2C_{N-2}^1\frac 1{N-1}p\left( 1-\frac 1{N-1}p\right)
^{N-3}$, and 
\[
\frac{dq_k}{dt}=-q_k+\frac Kb\left( q_{k-1}+q_{k+1}+M\right) . 
\]

$\cdots \cdots $.

($1.n$). There is $n$ random bonds on $\sigma _k$, and the probability is $%
\left( 1-p\right) ^2C_{N-2}^n\left( \frac 1{N-1}p\right) ^n\left( 1-\frac 1{%
N-1}p\right) ^{N-n-2}$. This group can be further divided into $C_{N-1}^n$
subgroups, each corresponding to $n$ specific spins connected to $\sigma _k$
via the random bonds, and with the same probability $\left( 1-p\right)
^2\left( \frac 1{N-1}p\right) ^n\left( 1-\frac 1{N-1}p\right) ^{N-n-2}$. In
the $i$th subgroup, 
\[
\frac{dq_k}{dt}=-q_k+\frac Kb\left(
q_{k-1}+q_{k+1}+\sum_{l=1}^nq_{i_l}\right) . 
\]
Thus averaging them we get, 
\[
\frac{dq_k}{dt}=-q_k+\frac Kb\left( q_{k-1}+q_{k+1}+nM\right) . 
\]

$\cdots \cdots $.

($1.(N-2)$). There is $N-2$ random bonds on $\sigma _k$, and the probability
is $\left( 1-p\right) ^2\left( \frac 1{N-1}p\right) ^{N-2}.$%
\begin{eqnarray*}
\frac{dq_k}{dt} &=&-q_k+\frac Kb\left(
q_{k-1}+q_{k+1}+\sum_{l=1}^{N-2}q_{i_l}\right) \\
&=&-q_k+\frac Kb\left( q_{k-1}+q_{k+1}+\left( N-2\right) M\right) .
\end{eqnarray*}
Thus, the first part of the time derivative of single-spin, 
\begin{eqnarray}
\left( \frac{dq_k}{dt}\right) _1 &=&\left( 1-p\right) ^2\left[ -q_k+\frac Kb%
\left( q_{k-1}+q_{k+1}\right) \right.  \nonumber \\
&&\left. +\frac Kb\sum_{n=0}^{N-2}C_{N-2}^n\left( \frac p{N-1}\right)
^n\left( 1-\frac 1{N-1}p\right) ^{N-n-2}nM\right] .
\end{eqnarray}

($2$). The bond connecting $\sigma _{k-1}$ and $\sigma _k$ is left unchanged
but that connecting $\sigma _k$ and $\sigma _{k+1}$ has been rewired. This
bond can be rewired to each one of the $N-1$ spins with equal probability $%
\left( 1-p\right) p/N$. In each case, we can analyze the situation in the
way described above. For example, the bond is rewired to $\sigma _j.$

($2.j.0$). There are no random bond on $\sigma _k$. The probability is $%
\left( 1-p\right) \frac pN\left( 1-p\right) ^2\left( 1-\frac 1{N-1}p\right)
^{N-2}$. 
\[
\frac{dq_k}{dt}=-q_k+\frac Kb\left( q_{k-1}+q_j\right) 
\]

($2.j.1$). There is only one random bond on $\sigma _k$. The probability is $%
\left( 1-p\right) \frac pNC_{N-2}^1\frac 1{N-1}p\left( 1-\frac 1{N-1}%
p\right) ^{N-3}.$%
\[
\frac{dq_k}{dt}=-q_k+\frac Kb\left( q_{k-1}+q_j+M\right) . 
\]

$\cdots \cdots $.

($2.j.n$). There is $n$ random bonds on $\sigma _k$, and the probability is $%
\left( 1-p\right) \frac pNC_{N-2}^n\left( \frac 1{N-1}p\right) ^n\left( 1-%
\frac 1{N-1}p\right) ^{N-n-2}$. 
\[
\frac{dq_k}{dt}=-q_k+\frac Kb\left( q_{k-1}+q_j+nM\right) . 
\]

$\cdots \cdots $.

($2.j.(N-2)$). There is $N-2$ random bonds on $\sigma _k$, and the
probability is $\left( \frac 1{N-1}p\right) ^{N-2}.$%
\[
\frac d{dt}q_k=-q_k+\frac Kb\left( q_{k-1}+q_j+\left( N-2\right) M\right) . 
\]
Thus the second part of the time derivative of single-spin, 
\begin{eqnarray}
\left( \frac{dq_k}{dt}\right) _2 &=&\left( 1-p\right) p\left[ -q_k+\frac Kb%
\left( q_{k-1}+M\right) \right.  \nonumber \\
&&\left. +\frac Kb\sum_{n=0}^{N-2}C_{N-2}^n\left( \frac p{N-1}\right)
^n\left( 1-\frac 1{N-1}p\right) ^{N-n-2}nM\right] .
\end{eqnarray}

($3$). The bond connecting $\sigma _k$ and $\sigma _{k+1}$ is left unchanged
but that connecting $\sigma _{k-1}$ and $\sigma _k$ has been rewired (we
omit the very small probability that this bond may be ''rewired'' to $\sigma
_k$). Now there is only one regular bond on $\sigma _k$. (Pay attention that
this case is different from ($2$).) Based on similar consideration, we have
the third part of the time derivative of single-spin obeys 
\begin{eqnarray}
\left( \frac{dq_k}{dt}\right) _3 &=&\left( 1-p\right) p\left[ -q_k+\frac Kb%
q_{k+1}\right.  \nonumber \\
&&\left. +\frac Kb\sum_{n=0}^{N-2}C_{N-2}^n\left( \frac p{N-1}\right)
^n\left( 1-\frac 1{N-1}p\right) ^{N-n-2}nM\right] .
\end{eqnarray}

($4$). Both the bond connecting $\sigma _k\sigma _{k+1}$ and $\sigma
_{k-1}\sigma _k$ have been rewired. Based on similar consideration, we have
the fourth part of the time derivative of single-spin 
\begin{equation}
\left( \frac{dq_k}{dt}\right) _4=p^2\left[ -q_k+\frac KbM+\frac Kb%
\sum_{n=0}^{N-1}C_{N-2}^n\left( \frac p{N-1}\right) ^n\left( 1-\frac 1{N-1}%
p\right) ^{N-n-2}nM\right] .
\end{equation}

Applying Eq. (\ref{important-relationship}) we get 
\begin{equation}
\frac{dq_k}{dt}=\sum_{i=1}^4\left( \frac{dq_k}{dt}\right) _i=-q_k+p\frac{2K}b%
M+\left( 1-p\right) \frac Kb\left( q_{k-1}+q_{k+1}\right) .
\label{result-app}
\end{equation}

\null\vskip0.2cm

\centerline{\bf Caption of figures} \vskip1cm

Fig.1. The $n$-dependence of critical point $K_c$ of $1$D Ising model on
A-SWN and R-SWN.

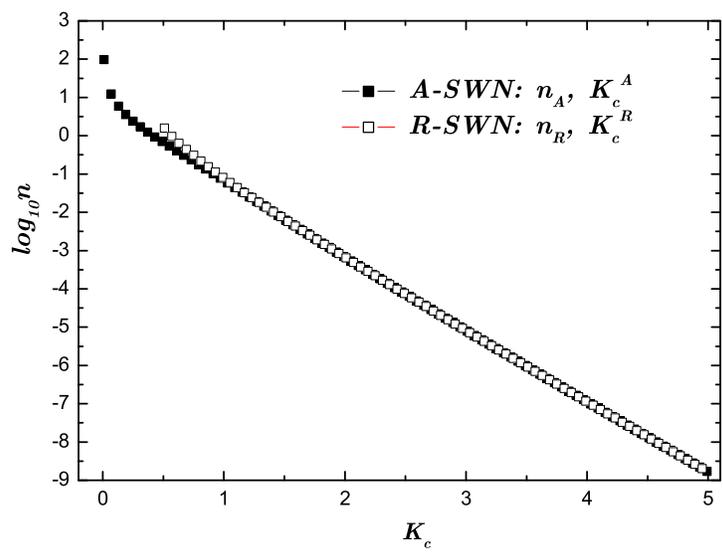
\end{document}